\RequirePackage{fix-cm}
\PassOptionsToPackage{numbers,sort&compress}{natbib}
\documentclass[smallextended, envcountsect, natbib]{svjour3}       %
\smartqed  %
\usepackage{graphicx}

\usepackage[hidelinks]{hyperref}
\usepackage{booktabs}
\usepackage{amsmath}

\usepackage{subfig}
\usepackage{textcmds}
\usepackage{cleveref}
\usepackage{caption}

\usepackage{enumitem}
\setlistdepth{9}

\usepackage{textcomp}
\usepackage{etoolbox}

\setlist[itemize,1]{label=$\bullet$}
\setlist[itemize,2]{label=$\bullet$}
\setlist[itemize,3]{label=$\bullet$}
\setlist[itemize,4]{label=$\bullet$}
\setlist[itemize,5]{label=$\bullet$}
\setlist[itemize,6]{label=$\bullet$}
\setlist[itemize,7]{label=$\bullet$}
\setlist[itemize,8]{label=$\bullet$}
\setlist[itemize,9]{label=$\bullet$}

\renewlist{itemize}{itemize}{9}

\usepackage{custom_macros}

\usepackage{xcolor}

\begin{document}

\title{Patterns of Multi-Container Composition for Service Orchestration with Docker Compose%
}

\titlerunning{Patterns of Multi-Container Composition}        %

\author{Kalvin Eng \and
        Abram Hindle \and 
        Eleni Stroulia %
}

\authorrunning{Eng et al.} %

\institute{Kalvin Eng \and
        Abram Hindle \and 
        Eleni Stroulia \at
              University of Alberta \\
              Edmonton, Alberta, Canada\\
              \email{\{kalvin.eng, abram.hindle, stroulia\}@ualberta.ca}           %
}

\date{Received: date / Accepted: date}

\maketitle

\begin{abstract}
Software design patterns present general code solutions to common software design problems. Modern software systems rely heavily on containers for running their constituent service components. Yet, despite the prevalence of ready-to-use Docker service images ready to participate in multi-container service compositions of applications, developers do not have much guidance on how to compose their own Docker service orchestrations. Thus in this work, we curate a dataset of successful projects that employ Docker Compose as an orchestration tool to run multiple service containers; then, we engage in qualitative and quantitative analysis of Docker Compose configurations. The collection of data and analysis enables the identification and naming of repeating multi-container composition patterns that are used in numerous successful open-source projects, much like software design patterns. These patterns highlight how software systems are orchestrated in the real-world and can give examples to anybody wishing to compose their own service orchestrations. These contributions also advance empirical research in software engineering patterns as evidence is provided about how Docker Compose is used.
\end{abstract}

\section{Introduction}

Following the successful adoption of software design patterns~\citep{Gamma1994-rf}, we analyze the patterns of successful self-deployed open-source projects employing Docker Compose as an orchestration tool to run multiple service containers. Software design patterns~\citep{Gamma1994-rf} provide solutions to recurring problems in software development, often at the class level, sometimes intersecting with architecture. %

Software systems have become more diverse over time as monolithic architectures and n-tier/n-layered projects transition to containerized deployments.  now require developers to be familiar with many tools and services of a software system that are often rapidly changing~\citep{carey_2021}. This diversity motivates the need to identify common composition patterns of services that could be reused by projects to improve simplicity. Since many projects use off-the-shelf container images to orchestrate services with different configuration parameters, we can identify composition patterns, such as using a reverse proxy service to route to services hosted internally.

In this work, we curate a dataset of successful projects that employ Docker Compose as a service orchestration tool from the self-hosted GitHub community; then, we engage in qualitative and quantitative analysis of Docker Compose configurations. Successful projects are those that have captured the attention and interest of a wider community, thus being recognized as valuable and significant. The collection of data and analysis enables the identification and naming of repeated composition patterns orchestrated by numerous successful open-source projects, much like software design patterns.

Our work derives  patterns from the open coding of Docker Compose multi-container compositions in projects of the self-hosted GitHub community. We identify three dimensions of Docker Compose usage from open coding: the service level, the orchestration level, and the repository level. The service level consists of the images and instructions used in each service of a Docker Compose file. The orchestration level refers to how the services interact among each other in a Docker Compose file when they are orchestrated together. The repository level examines how Docker Compose files are executed among other files in a repository. These dimensions categorize the observed patterns of Docker Compose usage. Patterns are considered to be commonly observed solutions that can address specific problems when using Docker Compose. 

By observing these patterns, developers can use them as a point of reference when deciding how to compose their services for orchestration. These patterns are beneficial as Docker Compose's versatility can make configurations unnecessarily complex~\citep{docker-versatility}. Furthermore, previous research~\citep{ibrahim2021study} fails to capture the essence of usage for Docker Compose as data is analyzed at a macro level without considering the minute details of service orchestration usage, such as how developers can use Docker Compose in different contexts like development and testing.

Our work is driven by the desire to allow practitioners to see real-world examples of Docker Compose patterns that encompass multi-container composition. Therefore, this work contributes the following:
\begin{itemize}
    \item An open-coded dataset of concepts in Docker Compose files;
    \item A set of Docker Compose patterns for software systems.
\end{itemize}
These contributions are helpful for anybody who uses Docker Compose to define their multi-container compositions for service orchestration and needs examples of . Furthermore, these contributions advance empirical research in software engineering patterns based on how Docker Compose is used.

\section{Background and Related Work}

In this section, we outline what multi-container composition and service orchestration is, how service containers are composed and orchestrated in the Docker ecosystem, and introduce related work motivating why we wish to investigate patterns in Docker Compose files.

\subsection{Multi-Container Composition and Service Orchestration}
Our use of the words: \textit{composition} and \textit{orchestration} stems from their usage in service-oriented architectures (SoA) in the past. Composition refers to a \qq{composite relationship between a collection of services}~\citep{erl2005service}. While orchestration refers to \qq{a centrally controlled set of workflow logic facilitates interoperability between two or more different applications}~\citep{erl2005service}.

Using the definition of composition from SoA, we define \textit{multi-container composition} as a collection of service containers that are associated with each other.

Orchestration has evolved from its SoA definition to be a term that is inferred as \textit{container orchestration} at the time of this paper writing in .
Container orchestration is a term that has been widely attributed to Kubernetes where multiple application containers are managed across different host machines~\citep{k8s}. The common definition for \qq{container orchestration} has appeared in survey literature as the deployment of containerized applications across computer clusters~\citep{rodriguez2019container, truyen2019comprehensive}. The Docker alternative to Kubernetes is Docker Swarm, which uses the Docker Compose file format to define the services, networks, and volumes of the containers~\citep{container-orchestrator}.

Docker Compose falls short of being a complete container orchestration solution as it solely manages and deploys containers on a single host instead of a cluster of host machines~\citep{container-orchestrator}. However, Docker Compose encompasses the aspect of managing multiple containers and is considered to be an intermediate solution between running a single container and running multiple containers across multiple hosts~\citep{miell2019docker, kane2023docker, container-orchestrator}. Docker Compose has been referred to as container orchestration by \citet{raj2015learning} and \citet{miell2019docker}. However, for the sake of clarity, we use in this paper the term \textit{service orchestration} to describe Docker Compose which manages and deploys multiple service containers locally on a host machine.

\subsection{The Docker Ecosystem}
Docker is a suite of \textit{Infrastructure as Code} (IaC) tools that can orchestrate containers for deployment on varying environments from a local machine to the cloud~\citep{guerriero2019adoption}. To orchestrate containers on a local machine, one would use the Docker Compose tool. While one would use the Docker Swarm tool to orchestrate containers across multiple machines in the cloud.

\begin{figure}[htbp]
\centering
  
\includegraphics[width=0.6\linewidth]{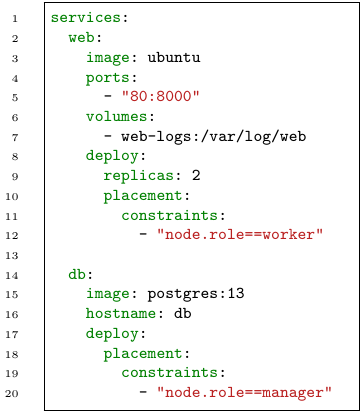}
\caption{Example of a Docker Compose file}
\label{docker-compose-example}
\end{figure}

Docker is designed to create \textit{images} and run \textit{instances} of these images called \textit{containers} which are defined in a Dockerfile. A Dockerfile is a set of instructions that are defined to create a containerized environment for running an application. In order to orchestrate multiple containers at once, Docker Compose or Docker Swarm can be used to facilitate the orchestration through the use of configurations defined via a YAML based Docker Compose file~\citep{ibrahim2021study}. The Docker Compose file describes a composition of services which can span multiple containers. An example can be seen in \Cref{docker-compose-example}.

\Cref{docker-compose-example} is an example of a simple Docker Compose file where the services \qq{web}, that is an Ubuntu container, and \qq{db}, that is a pre-built container image, are defined. At a minimum a Docker Compose file will always have at least one service defined with the option to give additional instructions such as mounting volumes, exposing ports, passing environment variables, and what to do in a swarm configuration. A swarm configuration allows for replicas of services to be deployed to a cluster which also uses Docker Compose configurations. As of , a Docker Compose file has 194 different keywords for configuration available , making it highly configurable for different use cases.

In 2020, the Docker Compose file format was released as open source under the name \qq{Compose Specification}~\citep{compose-spec-announce,compose-spec}. The \qq{Compose Specification} is used by alternative Docker Compose service orchestration tools such as nerdctl~\citep{nerdctl}, Okteto~\citep{okteto}, Podman~\citep{podman}, and Rancher~\citep{rancher}. The specification has also been used to help translate Docker Compose files to Kubernetes deployments with Kompose~\citep{kompose}.

Using Docker Compose files, a developer is quickly able to set up multiple containers for use cases such as development and testing. This makes the use of Docker a versatile choice in a software stack. However, the wide swath of instructions available in Docker Compose makes deployment complex as developers who use Docker may not be aware of possible features or configurations to use or to avoid. The flexibility of instructions makes learning Docker Compose more challenging as there are many options to solve a problem.  Therefore, this work seeks to identify patterns of configuration to better support the usage of Docker Compose in terms of simplicity, readability, and reproducibility.

\subsection{Previous Docker Studies and Prior Patterns Research}
\label{subsec:prior-work}
Since we wish to investigate patterns in Docker Compose files of the Docker ecosystem, we review prior work on understanding the Docker ecosystem. \citet{zhang2018insight, zhang2019clustering} look at the evolution of Dockerfiles to understand container build time and how images change in projects. 
They find that different Dockerfile architecture attributes, such as the number of image layers, the size of each layer, and the diversity of instructions, can affect the quality of Dockerfiles with regard to linter (static code analysis) errors and build time.
Furthermore, studies have thoroughly investigated Dockerfiles on public repository services and found that developers who build and maintain containers would benefit from utilizing tools for container creation~\citep{eng2021revisiting, lin2020large, cito2017empirical}. The empirical evidence observed in these studies supports the motivation for developing tools to improve the developer Docker experience, such as: DockerSlim~\footnote{\url{https://github.com/docker-slim/docker-slim}} which helps to optimize container images, dive~\footnote{\url{https://github.com/wagoodman/dive}} which helps to analyze container images, Buildpacks~\footnote{\url{https://buildpacks.io/}} which helps to build images, and Binci~\footnote{\url{https://github.com/binci/binci}} which helps containerize a development workflow. Despite the existence of tools aimed at enhancing the Docker experience, there still needs to be empirical research on patterns that can streamline the orchestration of various container contexts. This need motivates our exploration into Docker Compose configuration patterns.

Prior research has seen the potential in investigating service orchestration tools like Docker Compose.
\citet{xu2018mining} suggest that Docker Compose configurations can offer invaluable insights as multiple containers are \qq{glued} together to form a service highlighting why Docker Compose is important to investigate. \citet{ibrahim2021study} investigate the usage of Docker Compose and find that Docker Compose files are infrequently updated, advance Docker Compose operations are infrequently used, and that applications rarely adopt new versions of Docker Compose. Furthermore, they also find that over a quarter of Docker Compose configurations use a single container. This prior research motivates our work as it suggests that there is an interest into how to develop software systems with Docker Compose. 

Docker Compose can also be classified under \textit{Infrastructure as Code} (IaC), motivating the review of studies on IaC. In terms of related work for patterns in IaC, the most closely related work is by \citet{rahman2019seven} where they qualitatively identify security smells in 1,726 Puppet scripts using descriptive coding. From the descriptive coding, they describe and identify with qualitative analysis seven security smells that developers should be aware about. In a similar vein, \citet{ksontini2021refactorings} identify the refactoring types in Dockerfiles and Docker Compose configurations using manual analysis to suggest best practices in Docker. \citet{guerriero2019adoption} perform 44 semi-structured interviews with the findings that developers should find patterns for IaC that can be used across different IaC tools. This previous research suggest that a manual analysis of Docker Compose files could offer invaluable insights. 

Patterns have also been discovered by analyzing the commits and developers of repositories. \citet{rahman2020gang} look at Puppet file commits in repositories qualitatively to determine a taxonomy of defects. Furthermore, \citet{rahman2020code} also look at developers and their development activities and how some activities can be anti-patterns towards developing defect-free Puppet code. The discovery of patterns through analyses of development activity suggests that it can help promote more robust codebases.  

Patterns on other platforms have been investigated. \citet{shamim2020xi} perform a grey literature review of the best security practices for Kubernetes configurations. \citet{burns2016design} present cloud patterns observed in distributed systems using containers from anecdotal experiences. However, \citet{burns2016design} do not provide any empirical evidence of the cloud patterns used. The presentation of these patterns suggests that there is interest in understanding what patterns other developers are using.

From the Pattern Languages of Programs workshops, numerous general cloud patterns have been derived by the same group of authors: \citet{sousa2018overview, sousa2018engineering, sousa2018engineering-2, sousa2017engineering, sousa2015patterns}. These patterns are further investigated by the authors by performing a survey on 100 practitioners finding that practitioners use at least one of the patterns~\citep{sousa2021survey}. \citet{sousa2021survey} also find that 67\% of practitioners use containerization and 69\% of practitioners use container orchestration, and that the \qq{ease of use and security [are] main drivers for adoption} to the cloud which motivates the need for more patterns. Notably, none of the patterns by \citet{sousa2018overview, sousa2018engineering, sousa2018engineering-2, sousa2017engineering, sousa2015patterns}, to the best of our knowledge, have been presented in literature with empirical observations in real-world code.

Prior works on the Docker ecosystem have mainly focused on Dockerfile usage~\citep{eng2021revisiting, lin2020large, cito2017empirical, zhang2018insight, zhang2019clustering}. Research into patterns of IaC and development activities have not specifically addressed Docker Compose~\citep{rahman2019seven,ksontini2021refactorings,rahman2020gang,rahman2020code}. Therefore, our work focuses more on deriving the orchestration patterns of containers using Docker Compose similar to previous research for patterns in Infrastructure as Code. We elaborate on how the patterns we discover relate to the prior IaC works in \Cref{subsec:comparison}.

\section{Methodology}

Our methodology consists of three steps, separated into the subsections below. First, we select projects that use Docker Compose in the self-hosted GitHub community~\citep{awesome-selfhosted} and extract 527 files for analysis from 218 projects. Next, we use open coding to identify categories of Docker Compose file characteristics at a service level, orchestration level, and repository level. Finally, we describe patterns based on the categories extracted from open coding. The results of our analysis can be replicated using our replication package~\citep{replication-package}.

\subsection{Project Selection}
The data curated for this work was obtained by extracting GitHub repository URLs from the README of the \textit{Awesome-Selfhosted}~\citep{awesome-selfhosted} GitHub repository. The software projects of the self-hosted community often replace \textit{Software as a service} (SaaS) provider services with a diverse set of self-hosted solutions ranging from analytics software to wiki software.

We choose the list of self-hosted repositories to ensure we are studying projects that people use and deploy as the list is manually curated by the self-hosted community. The intended purpose of a repository --- which surrounds many of the perils of mining GitHub~\citep{kalliamvakou2014promises} --- might introduce noise into an analysis as repositories may not necessarily be used for the purposes of maintaining and writing code. A representative sample of typical Docker Compose usage in software systems can be shown by selecting a list curated by the self-hosted community, which frequently uses Docker Compose to launch software systems. This is beneficial because we are looking for deployable and real examples of patterns in Docker Compose files.

\begin{figure}[tbp]
\centering
\includegraphics[width=0.95\linewidth]{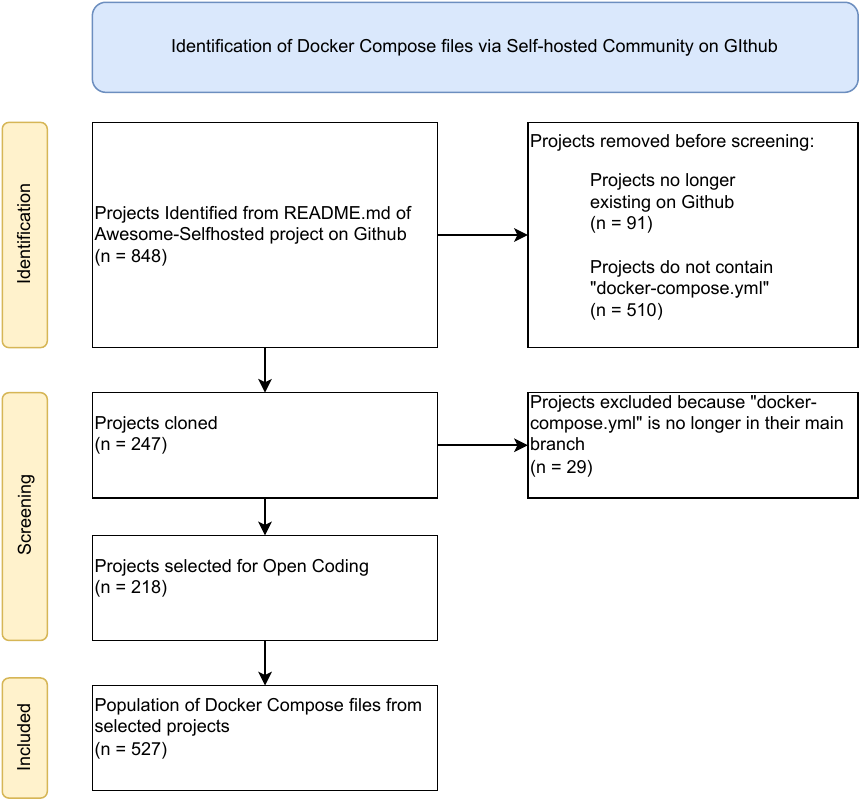}
  \caption{Project Selection Process}
\label{project-selection-process}
\end{figure}

Our project selection process  visualized in \Cref{project-selection-process}. Initially, 848 GitHub projects were extracted from the README pushed in the \qq{1726a40} commit. To determine if projects contain at least 1 Docker Compose file, we clone each of the repositories using the mirror option and the git history was queried across all refs and the HEAD for filenames containing \qq{docker-compose} (case-insensitive). From cloning, we obtain the actual commits and blob contents relating to the repositories for each of the repositories. The metadata was extracted using the PyDriller library~\citep{spadini2018pydriller}.  were able to obtain 247 successfully cloned projects  as of May 9, 2022.

\begin{table}[htbp]
\caption{Project Categories and Counts. Categories that have a count of 1 or 2 have been merged into Other~\cref{projects-count1}\textsuperscript{,}\cref{projects-count2}. \label{project_counts}}
\centering
\begin{tabular}{lr}
\toprule
Category &  Count \\
\midrule
Communication - Social Networks and Forums                                  &     12 \\
Communication - Custom Communication Systems                                &     11 \\
Automation                                                                  &     10 \\
E-commerce                                                                  &      9 \\
Bookmarks and Link Sharing                                                  &      8 \\
Feed Readers                                                                &      8 \\
Money, Budgeting \& Management                                               &      8 \\
Content Management Systems (CMS)                                            &      7 \\
Miscellaneous                                                               &      7 \\
Media Streaming - Video Streaming                                           &      6 \\
Task Management \& To-do Lists                                               &      6 \\
Communication - Email - Complete Solutions                                  &      5 \\
File Transfer - Single-click \& Drag-n-drop Upload                           &      5 \\
Note-taking \& Editors                                                       &      5 \\
Pastebins                                                                   &      5 \\
Personal Dashboards                                                         &      5 \\
Software Development - IDE \& Tools                                          &      5 \\
Software Development - Project Management                                   &      5 \\
Archiving and Digital Preservation (DP)                                     &      4 \\
Password Managers                                                           &      4 \\
Photo and Video Galleries                                                   &      4 \\
Software Development - API Management                                       &      4 \\
Wikis                                                                       &      4 \\
Conference Management                                                       &      3 \\
File Transfer - Web-based File Managers                                     &      3 \\
Internet of Things (IoT)                                                    &      3 \\
Learning and Courses                                                        &      3 \\
Media Streaming - Audio Streaming                                           &      3 \\
Polls and Events                                                            &      3 \\
Proxy                                                                       &      3 \\
Ticketing                                                                   &      3 \\
URL Shorteners                                                              &      3 \\
Other                                                                       &      28 \\
\bottomrule
\end{tabular}
\end{table}

Using the 247 cloned repositories, we find the latest main branch commit that has modified a file containing the name \qq{docker-compose}. We find that 29 projects have removed file(s) containing the name \qq{docker-compose} in their latest main branch commit. This indicates that Docker Compose may have been deprecated or moved elsewhere. Furthermore, we find that 99 projects have multiple files with the name \qq{docker-compose}. We remove the projects that no longer have \qq{docker-compose} in their latest main branch commit resulting in . To gain a better understanding of the diversity in the 218 projects, we outline in \Cref{project_counts} their respective categories as they appear in the \textit{Awesome-Selfhosted}~\citep{awesome-selfhosted} GitHub repository. In \Cref{project_counts} we combine categories with a count of 1
and a count of 2 into \qq{Other}~\footnote{\label{projects-count1}Projects with count of 1:
\textit{Booking and Scheduling}, \textit{Office Suites}, \textit{Search Engines}, \textit{Resource Planning}, \textit{Communication - SIP}, \textit{Human Resources Management (HRM)}, \textit{Maps and Global Positioning System (GPS)}, \textit{Document Management - Integrated Library Systems (ILS)}, \textit{Calendar \& Contacts - CalDAV or CardDAV Servers}, \textit{Read-it-later Lists}}\footnote{\label{projects-count2}Projects with count of 2:
\textit{Resource Planning - Enterprise Resource Planning}, 
\textit{Groupware}, 
\textit{Software Development - FaaS \& Serverless}, 
\textit{Gateways and Terminal Sharing}, 
\textit{File Transfer - Peer-to-peer Filesharing}, 
\textit{File Transfer - Object Storage \& File Servers}, 
\textit{Document Management}, 
\textit{Community-Supported Agriculture (CSA)}, 
\textit{Document Management - E-books},
\textit{Document Management - Institutional Repository and Digital Library Software},
\textit{Communication - Email - Mailing Lists and Newsletters}, 
\textit{Communication - Email - Webmail Clients}, 
\textit{Software Development - Localization}, 
\textit{Self-hosting Solutions}, 
\textit{Games}, 
\textit{Communication - XMPP - Servers}, 
\textit{File Transfer \& Synchronization}}.

From the 218 projects, we extract 527 Docker Compose-like files. The files are \qq{Docker Compose-like} because they might resemble a Docker Compose file, i.e., the file is YAML and is named \qq{docker-compose.yml} or co-exists with a file named \qq{docker-compose.yml}. 404 Docker Compose-like files are parts of projects that have more than one Docker Compose-like file, while 123 Docker Compose-like files belong to projects that have only a single Docker Compose-like file. We use these files for open coding as described in \Cref{sec:open-coding} below.

All the projects were cloned with considerations of the perils when curating git repositories highlighted by \citet{bird2009promises} such as: being aware of commits across branches; tracing changes that occur not only in the mainline branch; determining the branches of commits; and tracing the sources of merges. Therefore, in order to gain as much information is possible, all projects were cloned using the mirror option which is a total copy of the repository including all refs to track as many changes as possible. 

We also try to verify if these self-hosted projects represent products that developers might use by measuring the GitHub stars of each project. The measure of GitHub stars has been used in previous studies to measure the popularity of a project~\citep{brisson2020we, blincoe2016understanding, borges2016understanding} --- a project is assumed to be more popular if it has more stars, and as a result, by association, it should be used and important. %

\begin{figure}[tbp]%
    \centering
    \subfloat[\centering Self-Hosted projects\label{stars-self}]{{\includegraphics[width=.49\linewidth]{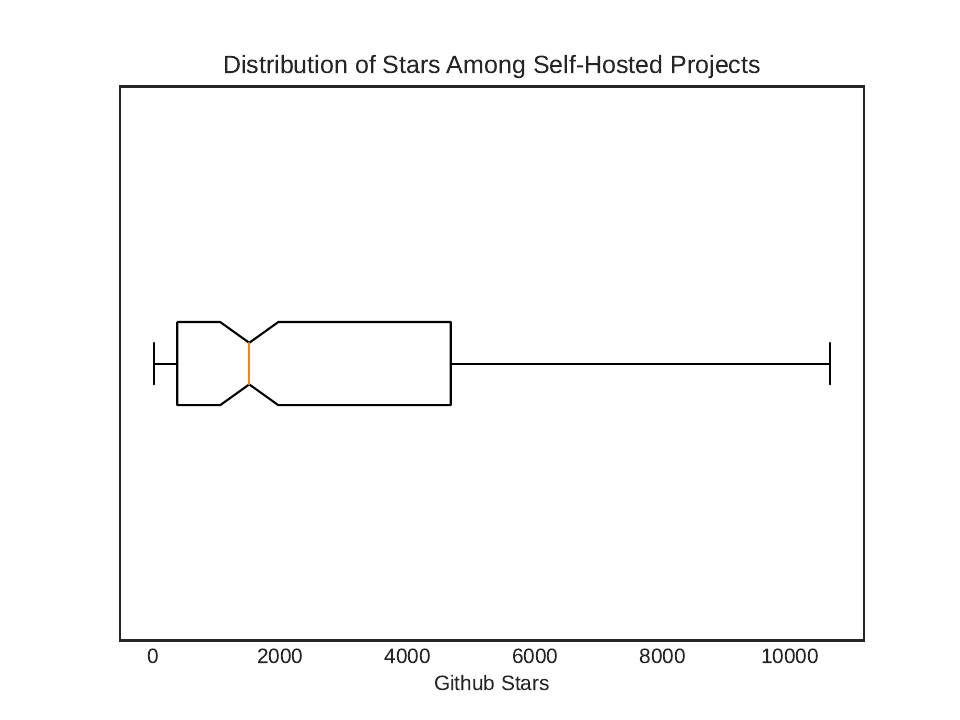} }}%
    \subfloat[\centering \citet{ibrahim2021study} Projects\label{stars-ibrahim}]{{\includegraphics[width=.49\linewidth]{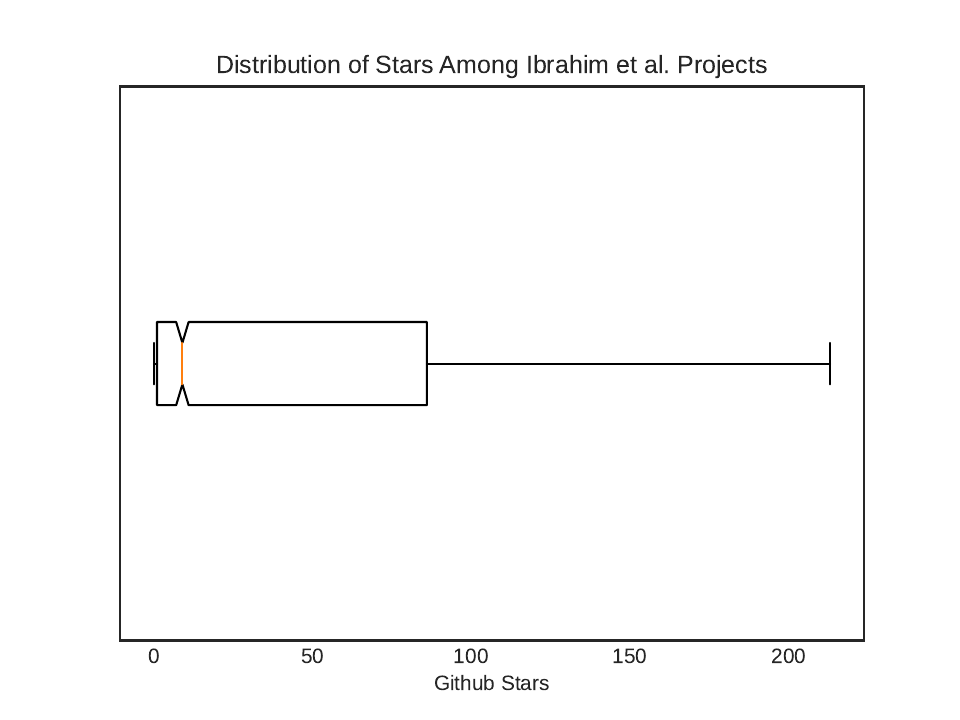} }}%
    \caption{Distribution of Stars}%
    \label{fig:stars}%
\end{figure}

To support why we should care about the selected self-hosted projects that we analyze, we count the GitHub stars --- the proxy for popularity  --- of each project as of December 20, 2022, as seen in the boxplot of \Cref{stars-self}. We compare the count of the stars in the self-hosted projects with the distribution of GitHub stars of Docker Compose projects selected by \citet{ibrahim2021study} from GitHub data in Google BigQuery seen in \Cref{stars-ibrahim}. . %

We use the Mann-Whitney U test, which indicates that the stars in the self-hosted projects are greater than the stars of projects selected by \citet{ibrahim2021study} with $p = 1.08\text{e-}94 < 0.05$. To ensure that the difference between the distributions are not only statistically significant but also practically significant, we get a large effect size from Cliff's Delta being $0.82 > 0.474$, which indicates that the greater stars in the selected self-hosted projects are practically significant. The greater number of stars demonstrates why the selected self-hosted projects are considered to be successful.

Overall, the star distributions are quite different as seen in \Cref{fig:stars}. The self-hosted projects' distribution has larger values, while \citepos{ibrahim2021study} distribution has smaller values suggesting that the sample of self-hosted projects  With a median of 1,512 and 25\% of the projects falling between 14 stars and 382 stars and 25\% having at least 4,674 stars, the self-hosted projects' have more stars overall. In comparison, \citepos{ibrahim2021study} distribution has a smaller number of stars overall, with a median of 9 and 25\% of the projects falling between 0 and 1 stars and 25\% of projects having at least 86 stars. Therefore, the higher number of stars in our chosen self-hosted projects is indicative of projects being used in practical applications compared to the prior Docker Compose usage study~\citep{ibrahim2021study}.

Notably, in the previous Docker Compose study~\citep{ibrahim2021study} %

\subsection{Open Coding}
\label{sec:open-coding}
\begin{figure}[tbp]
\centering
\includegraphics[width=1\linewidth]{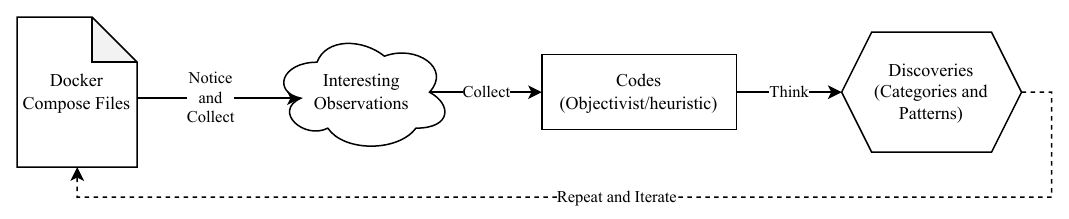}
  \caption{Open Coding Process Outline Inspired by \citet{seidel1998qualitative}}
\label{open-coding-process}
\end{figure}

Open coding is the process of finding concepts through manual inspection to gain deeper insights within data. \citet{seidel1998qualitative} describes it as a qualitative data analysis process where we \textit{notice}, \textit{collect} and \textit{think} about interesting things. The process is non-linear as observations can be re-iterated through many times. Qualitative methods are useful for coming up with observations that are difficult to derive with quantitative analysis. This process can be visualized in \Cref{open-coding-process}.  

With the collected Docker Compose files, a rater analyzes the files for what they consider to be \qq{interesting} observations and \qq{important} observations. \qq{Interesting} observations catch the attention of the rater and may appear infrequently while \qq{important} observations emerge from the repeated analysis of files from noticing common characteristics among the different files. Interesting observations within the files are \textit{noticed} and \textit{collected}. Any important observations are also \textit{noticed} and \textit{collected} separately with a label. 

\citet{seidel1998qualitative} says that labels of important observations can be \textit{thought} about and abstracted into \qq{in vivo codes} which are observations directly from the data or \qq{constructed codes} where observations are created by the annotator. From these codes a combination of \textit{objectivist} and \textit{heuristic} codes can emerge. \textit{Objectivist} codes occur when observations require some sort of interpretation as opposed to \textit{heuristic} codes which are rote observations. Codes can also be a mixture of \textit{objectivist} and \textit{heuristic} codes~\citep{seidel1998qualitative}.

From the codes and interesting thoughts, discoveries of \qq{categories} and \qq{patterns} can emerge by \textit{thinking} about relations, similarities, and dissimilarities among the codes in relation to the interesting thoughts. This process is iterative and repeated until the annotator deems the categories and patterns satisfactory.

We can define our open coding process in different levels of detail. We examine each Docker Compose file as part of our ad hoc process on three different levels: (1) what and how services are defined, (2) how the services holistically interact when orchestrated together, and (3) how the files function within a repository context.  From the coding, we have several codes that emerge and are used to derive patterns as described in \Cref{pattern-selection}.

For each of the three levels, there are guiding questions that emerge for our open coding such as:
\begin{enumerate}
    \item \textbf{Service} - How are the services defined and what are the services?
    \item \textbf{Orchestration} - How are the services similar and different among each other and how do they interact with each other?
    \item \textbf{Repository} - How does Docker Compose work in the repository with files and folders?
\end{enumerate}

The codes from these guiding questions help to develop the patterns that we present in this paper. From 218 GitHub projects, we obtain 527 Docker Compose-like files for open coding resulting in 80 codes with: 40 codes at the service level, 35 codes at an orchestration level, and 5 codes at a repository level.

\subsubsection{Service Level}
At a service level, we inspect each service configuration of a Docker Compose file and create new service type codes when an existing service purpose cannot be categorized under existing codes. To identify service types, we consider the image of a service. 

If the image is pulled from a registry, we look at the README in the container registry and read the README to determine if there is a purpose for the image that can be used to identify the type of a service. If the purpose cannot be identified, then the repository containing the Docker Compose file is inspected for READMEs to indicate which services may be present in the software system and can help identify the service type. If the service type cannot still be determined, then the string of the image (e.g.\ \qq{webapp:tag} of the line \qq{image: webapp:tag} in a Docker Compose file) is searched in Google to determine a purpose and identify a service type.

If the image is created from a Dockerfile rather than being pulled, we attempt to determine the purpose of the service by examining the Dockerfile and README of the base image if it is present in the container registry. We inspect the instructions following the base image to determine the service container image. For example, a base image could be an operating system and the following instructions in the Dockerfile are installing software packages using the base image. Then the purpose of the service can be determined based on the software packages installed and used to categorize the service under a service type. Failing to identify a purpose, then the repository containing the Docker Compose file is inspected for READMEs to indicate which services may be present in the software system and can help identify the service type. If the service type can still not be determined, then the string of the image (e.g.\ \qq{webapp:tag} of the line \qq{FROM image: webapp:tag}) is searched in Google to determine a purpose and identify a service type.

From the inspection of the service level, codes of service types emerge which are used to identify the purpose of a service in a Docker Compose file. We identify 40 different codes in total with 33 types with a count more than 1 as seen in \Cref{tab:service-types} and organize the types into 7 higher categories (Core, Networking, Reliability, Setup, Data, Scheduling, and Miscellaneous). 

\textit{Core services} refer to services that contain the central logic for a software system. In this category, there are 2 services:  the \textit{Frontend service} which are applications which connect with a database or backend; and the \textit{Backend service} which are APIs of applications that connect with a database.

While \textit{networking services} refer to services that help network other services. This includes: the \textit{Reverse proxy service} which routes web traffic to different servers with examples of services being as \texttt{Nginx} and \texttt{Traefik}; the \textit{Discovery service} which helps connect microservices together; the \textit{DNS service} which runs a DNS server; and the \textit{HTTP accelerator service} refers to services that can speed up HTTP requests with a cache such as  \texttt{Varnish}.

\textit{Reliability services} help maintain other services. There are numerous services in this category including: the \textit{Testing service} that aids in testing using tools such as \texttt{Selenium} which automates tests; the \textit{Container management service} which polls image registries for updates and automatically update the container image; the \textit{Tracing service} which refers to services that help manage logs of applications; and the \textit{Event monitoring service} which monitors and reports the performance of other services.

\textit{Setup services} are services that help initialize other services such as: the \textit{Database init service} which helps initialize a database by running scripts for a database such as loading data; and the \textit{Setup service} which helps initialize or set up an application.

Moreover, \textit{Data services} refer to services that manage the data of a software system. There are 5 services in this category including: 
the \textit{Database service} which are services that systematically store data in persistent storage such as \texttt{Postgres} and \texttt{Mongo}; the \textit{Caching service} that systematically stores data in memory and may not be persistent; the \textit{Database administration service} which acts as a GUI interface to databases; the \textit{Object storage service} that stores unstructured data organized by files and folders; and the \textit{Data streaming service} which refers to services that allow data to be delivered at high throughput such as \texttt{kafka}.

\textit{Scheduling services} are used to automatically perform tasks which include: the \textit{Job scheduling service} that schedules tasks to do on a server; the \textit{Cron service} that schedules tasks that are run on the bash command line using \texttt{cron}; and the \textit{Workflow service} that can help with automation of services.

Finally, \textit{Miscellaneous services} encompass a diverse range of offerings that do not fit into the above categories, including but not limited to: 
the \textit{Mail service} which are mail servers that receive emails via protocols like SMTP; 
the \textit{Search service} utilizing software such as \texttt{Solr} to build large indexes for improved search speed; 
the \textit{Identity service} managing multiple identities across services with software such as \texttt{KeyCloak}; 
the \textit{Visualization service} for data visualization with tools like \texttt{Kibana}; 
the \textit{Certificate service} for generating SSL certificates; 
the \textit{Zipping service} for compressing files into a zipped format; 
the \textit{Message broker service} for passing messages from one service to another; 
the \textit{Image recognition service} providing image recognition through API; 
the \textit{Chat service} with platforms such as Mattermost; 
the \textit{Office service} offering tools for text, spreadsheets, and presentations; 
the \textit{Linux utilities service} running Linux utilities like \texttt{busybox}; 
the \textit{Secrets service} for storing application secrets; 
and the \textit{Hello world service}, demonstrating the proper functioning of Docker.

\begin{table}[tbp]
\caption{Counts of Service Type Codes \label{tab:service-types}}
\centering
\begin{tabular}{lr}
\toprule
Service Type Code &  Count \\
\midrule
Database service             &    365 \\
Frontend service             &    343 \\
Backend service              &    276 \\
Caching service              &    111 \\
Testing service              &     98 \\
Reverse proxy service        &     92 \\
Mail service                 &     62 \\
Search service               &     37 \\
Database administration service  &     26 \\
Object storage service       &     23 \\
Identity service             &     17 \\
Visualization service        &     17 \\
Job scheduling service       &     15 \\
Container management service &     14 \\
Event monitoring service     &     14 \\
Certificate service          &     11 \\
Database init service        &     11 \\
Zipping service              &     11 \\
Message broker service       &      8 \\
DNS service                  &      8 \\
Tracing service              &      8 \\
Image recognition service    &      6 \\
Cron service                 &      5 \\
Chat service                 &      4 \\
Office service               &      4 \\
Setup service                &      3 \\
Workflow service             &      3 \\
Linux utilities service      &      3 \\
Secrets service              &      2 \\
HTTP accelerator service     &      2 \\
Hello world service          &      2 \\
Discovery service            &      2 \\
Data streaming service       &      2 \\
\bottomrule
\end{tabular}
\end{table}

\subsubsection{Orchestration Level}
When looking at Docker Compose files with multiple services, it is essential to consider individual services in isolation while also examining the relationships between them. We can gain insights into how services interact when deployed during orchestration by taking a holistic view of services within a Docker Compose file. To perform this examination, we develop some guiding questions, which includes:
\begin{itemize}
    \item Do any of the services share the same Docker Compose options? (e.g.\ using the same ENV file)
    \item How do the services share the same Docker Compose options? (e.g.\ using YAML aliases)
    \item Do any of the services use the \qq{links} or \qq{depends\_on} instructions to connect each other?
    \item Does one service expose a port and another service connects to that port with an environment variable?
    \item Does a service extend from another service? Does a service extend from another file?
\end{itemize}
It should be noted that the \qq{links} we refer to is the one used in Docker Compose files and is different from the legacy container links used by the Docker Engine.
To acquire more insights into the service relationships within a Docker Compose file, we iteratively refine these questions depending on interesting discoveries that emerge during our analysis. 

Based on the holistic view of services, codes of service relationships emerge and can form patterns of Docker Compose options that can be reused. We identify 35 codes at the orchestration level that can be seen in \Cref{tab:holistic-service}. The table explicitly lists services instead of defining connections at a meta-level as we wish to determine empirically how different kinds of services are commonly configured and related.

\begin{table}[tbp]
\caption{Counts of Codes at a Orchestration Level \label{tab:holistic-service}}
\centering
\begin{tabular}{lr}
\toprule
Code &  Count \\
\midrule
Frontend service connects to database service with environment vars          &     28 \\
Duplicate image reuse with different commands                                &     17 \\
Frontend environment variables map to database service environment variables &     17 \\
Uses yaml aliases                                                            &     15 \\
Service labels are configuration for reverse proxy                           &     15 \\
Backend service connects to database service with env variable               &     12 \\
Uses \textit{extends} for additional instructions                                   &     11 \\
Frontend service connects to caching service with environment var            &     11 \\
Reverse proxy service shares volume with frontend service                    &      8 \\
Backend service connects to caching service with env variable                &      8 \\
Frontend shares volumes with backend service                                 &      4 \\
Certificate service shares volume with reverse proxy service                 &      3 \\
Backend services shares the same env file with the database service          &      3 \\
Database management service connects to database service with env variables  &      3 \\
Duplicate image reuse                                                        &      2 \\
Frontend connects to mail service with env var                               &      2 \\
Frontend service shares variable with caching service                        &      2 \\
Frontend service connects to backend service with environment var           &      2 \\
Cluster environment variables                                                &      1 \\
Duplicate image reuse with different env vars                                &      1 \\
Frontend services shares the same env file with the caching service          &      1 \\
Frontend services shares the same env file with the database service         &      1 \\
Reverse proxy service shares volume with backend service                     &      1 \\
Container management service shares volume with frontend service             &      1 \\
Frontend service shares variable with secrets service                        &      1 \\
Frontend service connects to message broker service with env variable        &      1 \\
Frontend service connects to object storage service with env variable        &      1 \\
Backend service connects to object storage service with env variable         &      1 \\
Backend service connects to search service with env variable                 &      1 \\
Reverse proxy maps to frontend service with env variable                     &      1 \\
Testing service connects to database service via env                         &      1 \\
Frontend service connects to search service with environment var            &      1 \\
Docker swarm instructions                                                    &      1 \\
Backend service shows relationship to identity service with env var          &      1 \\
Job scheduling service connects to database service with env war             &      1 \\
\bottomrule
\end{tabular}
\end{table}

\subsubsection{Repository Level}
In a repository context, we look at how Docker Compose files coexist among other files and folders in the repository.
In addition to the Docker Compose files themselves, we carefully examine other relevant files to gain a more complete understanding of how they are used in practice. These files include READMEs, which often provide valuable information on the intended usage and context of a Docker Compose file. Moreover, we also analyze the configurations associated with these files, such as ENV files which can indicate that a Docker Compose file is used for development when the debug flag is set. Execution details of Docker Compose can also be found by looking at Makefiles and bash scripts. In addition to looking at the contents of files, the filenames and file paths are also inspected, i.e., we look for keywords such as \qq{development} and \qq{testing} to indicate a Docker Compose file purpose.

The repository context helps identify the codes of use cases for Docker Compose files and codes of how Docker Compose files are executed. At a repository context level, we identify 5 codes as seen in \Cref{tab:repository-context}.

\begin{table}[htb]
\caption{Counts of Codes at a Repository Level \label{tab:repository-context}}
\centering
\begin{tabular}{lr}
\toprule
Code &  Count \\
\midrule
Override file                                              &     53 \\
Auto-generated file                                        &     29 \\
Not an actual docker compose file, configuration           &     21 \\
Not an actual docker compose file, template for generating &      2 \\
Development use case from comment                          &      2 \\
\bottomrule
\end{tabular}
\end{table}

The codes in \Cref{tab:repository-context} can be described as follows:
\begin{itemize}
    \item \qq{Override file} refers to Docker Compose files that contain service configurations that can override other configurations when executed with another Docker Compose file.
    \item \qq{Auto-generated file} are Docker Compose files that have been automatically generated using a tool.
    \item \qq{Not an actual docker compose file, configuration} means that files appear to be Docker Compose-like but are actually meant to configure services.
    \item \qq{Not an actual docker compose file, template for generating} are files that are a template for generating Docker Compose files.
    \item \qq{Development use case from comment} refers to the use case of a Docker Compose file that is determined to be for the development environment and testing based on comments in the Docker Compose file.
    \item \qq{Cluster environment variables} are environment variables that define how to connect to other instances of a service.
\end{itemize}

\subsection{Data for Pattern Discovery}
\label{pattern-selection}
Through the open coding, we identify patterns that are common or interesting to developers and report ones that occur more than once in \Cref{sec:patterns}. These patterns stem from interesting observations that we note down and describe in \Cref{subsec:interesting-patterns}. We also quantitatively identify patterns by doing the following among the levels of open coding:
\begin{itemize}
    \item Frequent itemset mining of the service type codes as described in \Cref{subsec:frequent-itemset}.
    \item Considering the counts of orchestration level codes as described in \Cref{subsec:obs-counts}.
\end{itemize}

\subsubsection{Interesting Observations}
\label{subsec:interesting-patterns}
Observations are considered \qq{interesting} when certain attributes of a Docker Compose file are considered by the reviewer to be unusual among all other Docker Compose files. For our interesting observations we note the following:
\begin{itemize}
    \item The \texttt{mailu}~\footnote{https://github.com/Mailu/Mailu/tree/e8641245} project generates Docker Compose from templates (they could have possibly used overrides) and there's other kinds of Docker Compose files in its repo;
    \item Many projects (e.g. \texttt{httplaceholder} \footnote{\url{https://github.com/dukeofharen/httplaceholder/tree/74d0b34/docker}}) host multiple example Docker Compose files to demonstrate how to run their application with different service stacks;
    \item Projects using override files are usually highly configurable and are compatible with different service stacks;
    \item The \texttt{lardbit/nefarious}~\footnote{\url{https://github.com/lardbit/nefarious/tree/a6f49a1}} project could have used YAML aliases instead of extends;
    \item There is an override file to change from pulling an image to building an image in the \texttt{mailcow} project;
    \item YAML aliases are used to repeat services with the same images but different commands;
    \item The \texttt{piqueserver}~\footnote{\url{https://github.com/piqueserver/piqueserver/tree/b30dab7d}} project \qq{reverses} override usage where an image is defined in the override file, and the base template is empty;
    \item Docker Compose has services that do not pertain to project stack directly (e.g.\ Using letsencrypt pebble);
    \item In a Docker Compose file, sometimes the \qq{extends} configuration filename does not contain the string \qq{docker-compose} when it typically does --- this can be seen in the \texttt{rero-ils}~\footnote{\url{https://github.com/rero/rero-ils/tree/9fcf2e9e}} project;
    \item Docker Compose files for each microservice are automatically generated in the \texttt{overleaf}~\footnote{\url{https://github.com/overleaf/overleaf/tree/ed66b43/services}} project. 
\end{itemize}

\noindent %

\subsubsection{Frequent Itemset Mining}
\label{subsec:frequent-itemset}
Frequent itemset mining of the service types helps find sets of services that frequently co-occur together. 
We only consider service type codes as they are easily understood to be defined together as opposed to the other orchestration and repository codes that may not intuitively co-occur.
These frequent itemsets can signify a pattern of interest. To filter the amount of itemsets of interest, we consider itemsets that have support of at least 0.05 meaning that at least five percent of the Docker Compose files contain these services and are considered to be frequent itemsets. The results of our mining can be seen in \Cref{tab:freq-itemsets}.

\begin{table}[tbp]
\caption{Frequent Itemsets of Service Types \label{tab:freq-itemsets}}
\centering
\begin{tabular}{lr}
\toprule
 Support &                                                    Itemsets \\
\midrule
0.051233 &                         \textbf{(testing service, database service)} \\
0.053131 &                    (caching service, reverse proxy service) \\
0.055028 &        \textbf{(backend service, caching service, database service)} \\
0.056926 & \textbf{(reverse proxy service, frontend service, database service)} \\
0.060721 &                                              (mail service) \\
0.062619 &  \textbf{(backend service, reverse proxy service, database service)} \\
0.064516 &                                            (search service) \\
0.072106 &                          (backend service, caching service) \\
0.075901 &       \textbf{(backend service, frontend service, database service)} \\
0.079696 &       \textbf{(caching service, frontend service, database service)} \\
0.085389 &                    (backend service, reverse proxy service) \\
0.092979 &                                           (testing service) \\
0.096774 &                   (reverse proxy service, frontend service) \\
0.102467 &                   (reverse proxy service, database service) \\
0.121442 &                         (backend service, frontend service) \\
0.125237 &                         (caching service, database service) \\
0.129032 &                         (caching service, frontend service) \\
0.161290 &                                     (reverse proxy service) \\
0.178368 &                         \textbf{(backend service, database service)} \\
0.197343 &                                           (caching service) \\
0.278937 &                                           (backend service) \\
0.309298 &                        \textbf{(frontend service, database service)} \\
0.529412 &                                          \textbf{(database service)} \\
0.571157 &                                          (frontend service) \\
\bottomrule
\end{tabular}
\end{table}

In \Cref{tab:freq-itemsets}, we can see that there is high support for services connecting to a database service which are highlighted in bold. This motivates the patterns in \Cref{pattern:serv-database}  and \Cref{pattern:serv-data-cache} . Also in \Cref{tab:freq-itemsets}, we can note that reverse proxy services are also very common which motivates the \qq{HTTP Reverse Proxy Service} pattern in \Cref{pattern:reverse-proxy}. Noting that there are some very frequent items within itemsets is a good indicator as to what patterns can be derived.

\subsubsection{Observation Counts}
\label{subsec:obs-counts}
At the orchestration level and repository level we use the counts of combinations of observations as seen in \Cref{tab:holistic-service} and \Cref{tab:repository-context}. These counts help corroborate some of our interesting observations in \Cref{subsec:interesting-patterns}.

An example where we used counts to motivate a pattern is the observation \qq{Auto-generated file} found in \Cref{tab:repository-context} with a count of 29 for identifying the \qq{Automatic Docker Compose File Generation} pattern in \Cref{pattern:auto}.

\section{Patterns}
\label{sec:patterns}
In this section, we present 14 interesting patterns of Docker Compose usage discovered from interesting observations, frequent itemset mining, and observation counts as described in \Cref{pattern-selection}. Although a pattern may come across as common to some readers, there has been no empirical evidence or prior work to support these claims. More details of how the presented patterns were identified and its supporting evidence is described in this section using %

\begin{itemize}
    \item \textbf{Motivation.} Why developers might use this pattern.
    \item \textbf{Context/Applicability.} The situations where using the pattern might be appropriate.
    \item \textbf{Description.} An explanation of how the pattern works.
    \item \textbf{Advantages.} Description of the benefits of using this pattern compared to other patterns.
    \item \textbf{Disadvantages.} Description of the possible negative impacts of using this pattern.
    \item \textbf{Potential Issues.} Issues to be aware of when using this pattern.
    \item \textbf{Real-world Examples.} Examples identified from the Docker Compose configuration files within self-hosted software projects.
    \item \textbf{Supporting Evidence.} How the pattern and counts of the real-world examples were discovered.
\end{itemize}

\noindent Along with the template, we also include a visualization of the pattern to better help understand how the pattern works at a high-level.

\subsection{Automatic Docker Compose File Generation}
\label{pattern:auto}

\begin{figure}[htbp]
\centering
\includegraphics[width=1\linewidth]{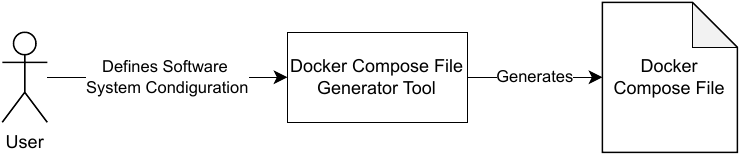}
  \caption{Pattern of Automatic Docker Compose File Generation.}
\label{fig-pattern:auto}
\end{figure}

\begin{itemize}
    \item \textbf{Motivation.} Software systems can be orchestrated in many different ways with different configurations and services. Due to the complexity of creating Docker Compose files with different configurations and services, tools have been created to automatically generate Docker Compose files as opposed to manually creating and editing Docker Compose files.
    \item \textbf{Context/Applicability.} Developers of software systems who wish to simplify their deployments for their software systems may wish to develop automatic Docker Compose generation tools. This leaves less ambiguity as to how their software system should be run and configured.
    \item \textbf{Description.} Docker Compose files are automatically generated using a tool that allows for the setting of the application specific variables and services that are needed to run the software system. The Docker Compose files generated can be run using one command without any other further configuration. %
    \item \textbf{Advantages.} As a user of the software system, the setup of the software systems is simplified as you can create the orchestration files with a guided tool. Docker Compose files are also consistent because of the automatic generation.
    \item \textbf{Potential Issues.} To recreate Docker Compose files, you will have to reuse the tool again to avoid any misconfigurations. Hand edits could lead to errors in deployment if modified incorrectly. The tool for automatic generation requires maintenance which can increase the dependencies and technical debt of a project.
    \item \textbf{Real-world Examples.} We found 29 Docker Compose files evidencing automatic Docker Compose file generation from web tools and build scripts. For example, the mail server project \texttt{Mailu} has created a web tool\footnote{\url{https://web.archive.org/web/20220915164516/https://setup.mailu.io/1.9/}} to generate Docker Compose files that can be run using Docker Compose or in a Docker Swarm. The \texttt{Overleaf} project\footnote{\url{https://github.com/overleaf/overleaf/tree/ed66b43/services}} also automatically generates continuous integration, development, and testing configurations for each of their microservices using build scripts.
    \item \textbf{Supporting Evidence.} These examples were found by counting the observations at a repository context level as seen in \Cref{tab:repository-context} denoted as \qq{Auto-generated file}. It is motivated by the interesting observation, \qq{Docker Compose files for each microservice are automatically generated in the \texttt{overleaf} project}.
\end{itemize}

\subsection{YAML Anchor and Alias}
\label{pattern:alias}

\begin{figure}[htbp]
\centering
\includegraphics[width=0.75\linewidth]{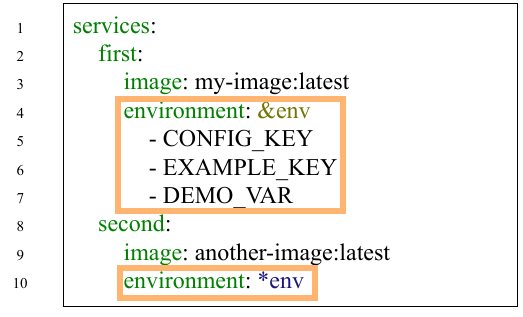}
  \caption{Example of how the YAML Anchor and Alias pattern is implemented from \citet{compose-alias}.}
\label{fig-pattern:alias}
\end{figure}

\begin{itemize}
    \item \textbf{Motivation.} The use of YAML anchors and aliases allow for the option of defining common sections of code in YAML files once so that it can be used repeatedly. This allows for Docker Compose file configurations to be neater and more efficient~\citep{compose-alias}.
    \item \textbf{Context/Applicability.} When services are re-used multiple times with similar configurations in the same Docker Compose file, YAML anchors and aliases can be used.
    \item \textbf{Description.} Docker Compose uses a YAML parser at its core, and therefore can use YAML and anchors aliases as defined in the YAML spec. An anchor \qq{\&} can be denoted for a section in a YAML file, where it is placed after a colon with a keyword as such \qq{\&keyword}. The section can then be repeated with an alias \qq{*} by specifying after a colon \qq{*keyword}. The repeated section can also be overwritten with \qq{$<<$} by specifying \qq{$<<$: *keyword} under a section. %
    \item \textbf{Advantages.} The use of YAML aliases prevents duplicate configuration code. 
    \item \textbf{Potential Issues.} The use of YAML aliases may appear non-intuitive at first due to the syntax. Furthermore, although it simplifies the Docker Compose file by less repetition, readability may still become an issue when there are many anchors and aliases to trace through.
    \item \textbf{Real-world Examples.} We found 15 Docker Compose files using this pattern. For example, the \texttt{chasiq} project\footnote{\url{https://github.com/chaskiq/chaskiq/blob/8fb33b/docker-compose.yml}} uses YAML anchors and aliases with overrides to configure their Docker Compose file. The canvas-lms project\footnote{\url{https://github.com/instructure/canvas-lms/blob/d4ee7b/docker-compose.yml}} uses YAML anchors and aliases with an override to allow their main application to be run as a job scheduler. The \texttt{minio} project also uses this pattern to repeat configurations\footnote{\url{https://github.com/minio/minio/blob/95d1a1/docs/orchestration/docker-compose/docker-compose.yaml}}.
    \item \textbf{Supporting Evidence.} The examples were found by counting the observations among services as seen in \Cref{tab:holistic-service} under \qq{Uses yaml aliases}.
    \item \textbf{Category.} %
    \item \textbf{Related Patterns.} %
\end{itemize}

\subsection{Docker Compose Service Inheritance}
\label{pattern:inheritance}

\begin{figure}[htbp]
\centering
\includegraphics[width=1\linewidth]{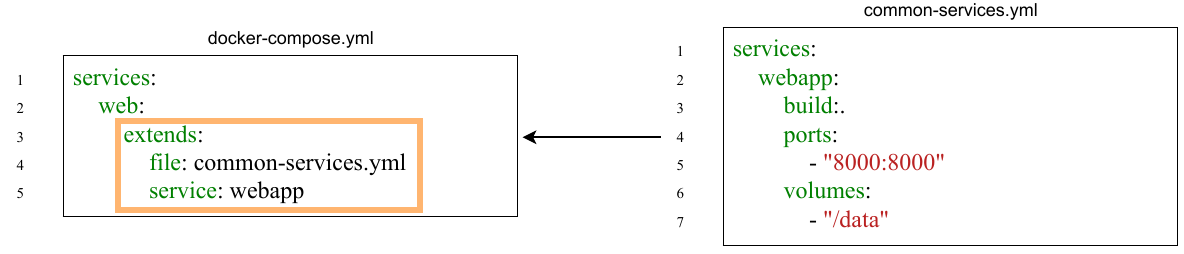}
  \caption{Example of how the Docker Compose Service Inheritance pattern is implemented from \citet{compose-extends}.}
\label{fig-pattern:inheritance}
\end{figure}

\begin{itemize}
    \item \textbf{Motivation.} Docker Compose service configurations are often similar with slight modifications. Therefore, the inheritance of configurations from other similar services can help simplify configurations by repeating services that are similar.
    As a feature, Docker Compose offers the \qq{extends} option which allows the inheritance of service configurations from other Docker Compose files at a service level or from services already within a Docker Compose file. %
    \item \textbf{Context/Applicability.} When services are re-used multiple times with similar configurations, the \qq{extends} option can be used to import service configurations from another Docker Compose file or existing services already in the Docker Compose file. This can allow for specialization of services within a Docker Compose file.
    \item \textbf{Description.} Under a defined service in a Docker Compose file, a service configuration can be imported by specifying the Docker Compose file path and the service that is to be extended. A service can also be extended by specifying a service name of the current file . The imported configuration can be overridden by re-specifying Docker Compose options. %
    \item \textbf{Advantages.} Prevents duplication of code.
    \item \textbf{Potential Issues.} Readability may become an issue as developers have to trace through Docker Compose files to find the imported configurations. Using extends on an existing service is not ideal when you do not want the base service to be orchestrated, to avoid orchestrating the base service the command line Docker Compose orchestration command needs to change.
    \item \textbf{Real-world Examples.} This pattern was discovered in 11 Docker Compose files. The \texttt{hawkpost} project\footnote{\url{https://github.com/whitesmith/hawkpost/blob/5d4f017/docker-compose.yml}} extends their services from a common configuration file and overwrites the common service configuration with service specific options. The \texttt{nefarious} project extends a service from within the same file\footnote{\url{https://github.com/lardbit/nefarious/blob/e8aa423/docker-compose.base.yml}} and from an external file\footnote{\url{https://github.com/lardbit/nefarious/blob/e8aa423/docker-compose.transmission-vpn.yml}}. The \texttt{mattermost} project\footnote{\url{https://github.com/mattermost/mattermost-server/blob/38d0c2/docker-compose.yaml}} also extends from an external file.
    \item \textbf{Supporting Evidence.} This pattern is found by counting the observation \qq{Uses \textit{extends} for additional instructions} as seen in \Cref{tab:repository-context} among the repository context.
    \item \textbf{Related Patterns.} %
\end{itemize}

\subsection{Docker Compose Override Use Case}
\label{pattern:override}

\begin{figure}[htbp]
\centering
\includegraphics[width=1\linewidth]{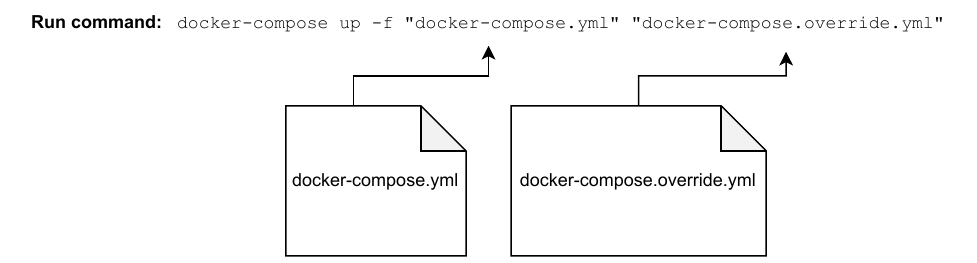}
  \caption{Example of how docker compose can override files in the command line.}
\label{fig-pattern:override}
\end{figure}

\begin{itemize}
    \item \textbf{Motivation.} A service may have different use cases whereby only small parts of a service configuration need to be changed. %
    \item \textbf{Context/Applicability.} When you need to run services in another mode such as testing and the Docker Compose file is designed for production deployment.
    \item \textbf{Description.} A different override Docker Compose file is created with the same services and configurations that they wish to override. This override Docker Compose file is called during Docker Compose orchestration in the command line with the \qq{-f} flag along with the main Docker Compose file. The result is that containers can be started up with overridden configurations of the main Docker Compose file. %
    \item \textbf{Advantages.} It separates use case scenarios of how Docker Compose is used. 
    \item \textbf{Potential Issues.} The command to run override configurations is not evident in the code base and requires explicit or inferred instructions of what to run. Override files may not also meet the minimum specifications for a Docker Compose to run, and will not run by itself.
    \item \textbf{Real-world Examples.} We found 53 examples of override files. The \texttt{share} project\footnote{\url{https://github.com/MrDemonWolf/share/blob/b231c/docker-compose.override.yml.example}} uses an override file to orchestrate a dev mode. The \texttt{board} project\footnote{\url{https://github.com/RestyaPlatform/board/blob/5804f/docker-compose.override.yml.example}} also uses an override file for development. The \texttt{otobo} project uses override files to change the ports of services with one being \qq{http.yml}\footnote{ \url{https://github.com/RotherOSS/otobo/blob/24d49/scripts/docker-compose/http.yml}}.
    \item \textbf{Supporting Evidence.} These examples of the pattern were found by counting the observations among repository context as seen in \Cref{tab:repository-context} under \qq{Override file}.
    \item \textbf{Related Patterns.} %
\end{itemize}

\subsection{Certificate Generation and Mapping}
\label{pattern:cert-gen}

\begin{figure}[htbp]
\centering
\includegraphics[width=0.5\linewidth]{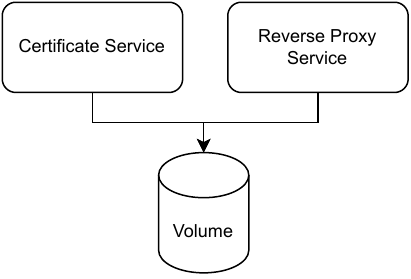}
  \caption{High-level overview of services in the Certificate Generation and Mapping pattern.}
\label{fig-pattern:cert-gen}
\end{figure}

\begin{itemize}
    \item \textbf{Motivation.} Developers wishing to provide secure transport (SSL/TLS) to their services may generate a certificate that is used through a reverse proxy.
    \item \textbf{Context/Applicability.} Certificates help enable encrypted communications between machines and also contain identification information about chains of authority and trust. Generating and managing these certificates can be cumbersome, hence the creation of container services to generate certificates for reverse proxy services.
    \item \textbf{Description.} A service container that generates certificates is used. The volumes containing the certificates of the certificate service are mapped to the same volumes of the reverse proxy service. The reverse proxy is configured to use the mapped volume certificates of the certificate service. %
    \item \textbf{Advantages.} The certificate issuance and management process is simplified via service orchestration as no certificate files need to be transferred and no command line operations need to be manually executed. 
    \item \textbf{Potential Issues.} It can be more difficult to debug between containers.
    \item \textbf{Real-world Examples.}  This pattern was discovered in 9 Docker Compose files. The \texttt{PeerTube} project uses certbot to issue a certificate and map it to the reverse proxy service with a volume.\footnote{\url{https://github.com/Chocobozzz/PeerTube/blob/1606ac2/support/docker/production/docker-compose.yml}} The \texttt{sish} project uses dnsrobocert to generate a certificate for its main application container.\footnote{\url{https://github.com/antoniomika/sish/blob/62035/deploy/docker-compose.yml}} The \texttt{azuracast} project generates a certificate and maps it to its reverse proxy service via a volume.\footnote{\url{https://github.com/AzuraCast/AzuraCast/blob/97aaa/docker-compose.multisite.yml}}
    \item \textbf{Supporting Evidence.} These examples were found by counting the Docker Compose file service types that both contained \textit{certificate service} and \textit{reverse proxy service} as seen in \Cref{tab:service-types}.
    \item \textbf{Related Patterns.} %
\end{itemize}

\subsection{Container Management Services}
\label{pattern:container-manage}

\begin{figure}[htbp]
\centering
\includegraphics[width=0.75\linewidth]{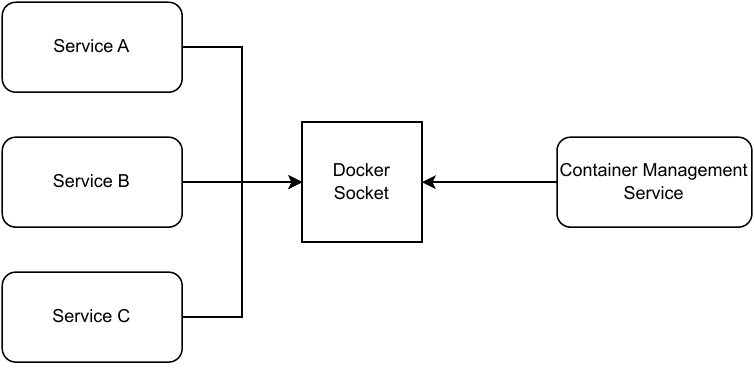}
  \caption{High-level overview of the Container Management Services pattern.}
\label{fig-pattern:container-manage}
\end{figure}

\begin{itemize}
    \item \textbf{Motivation.} To improve the deployment workflow, container management services are used to maintain the latest image versions of services in a container registry.
    \item \textbf{Context/Applicability.} Instead of re-pulling images and rebuilding the containers of your Docker Compose orchestration, a service can be started to poll a Docker Container registry for container image updates.
    \item \textbf{Description.} A container management service is set up to connect directly with the Docker socket and is configured to watch specific containers for updates in container registries. For example, the watchtower service polls the container registry for any updated to the associated image of a container and automatically updates the container if the associated image is updated. %
    \item \textbf{Advantages.} Using a container management service helps reduce maintenance costs as service containers can be upgraded to the latest version without any user intervention.
    \item \textbf{Potential Issues.} There is overhead to manage the management service. As well, since the container management service pulls automatically from a container registry, breaking changes to the container registry could break the current software system.
    \item \textbf{Real-world Examples.} We count 14 Docker Compose files using this pattern. The \texttt{budibase} project uses watchtower to manage container updates.\footnote{\url{https://github.com/Budibase/budibase/blob/56147/hosting/docker-compose.yaml}} The \texttt{egroupware} project also manages container updates with watchtower.\footnote{\url{https://github.com/EGroupware/egroupware/blob/1c053/doc/docker/docker-compose.yml}} While the \texttt{appsmith} project uses watchtower to update containers.\footnote{\url{https://github.com/appsmithorg/appsmith/blob/382ea53/app/server/appsmith-server/src/main/resources/docker-compose.yml}} 
    \item \textbf{Supporting Evidence.} The examples were found by counting the service types as seen in \Cref{tab:service-types} under \textit{Container management service}.
\end{itemize}

\subsection{Database Initialization Service with Database Service}
\label{pattern:data-init}

\begin{figure}[htbp]
\centering
\includegraphics[width=0.5\linewidth]{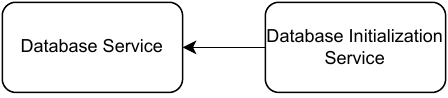}
  \caption{High-level overview of the Database Initialization Service with Database Service pattern.}
\label{fig-pattern:data-init}
\end{figure}

\begin{itemize}
    \item \textbf{Motivation.} Databases are initialized via an external service that shuts down after the task is run.
    \item \textbf{Context/Applicability.} To import data into an existing database, a database initialization service can be created.  This is often seen in development and testing service orchestrations.
    \item \textbf{Description.} The database initialization service is mapped to the database service via environment variables. The database initialization service runs scripts to initialize the database in the database service. %
    \item \textbf{Advantages.} Short running initialization container that is separate from the main application service allows for separation of concerns in the software system.
    \item \textbf{Potential Issues.} There is an overhead cost for setting up the initialization service where data must be dumped and scripts need to be created to import the data.
    \item \textbf{Real-world Examples.} We count 9 Docker Compose files using this pattern. The \texttt{budibase} project\footnote{\url{https://github.com/Budibase/budibase/blob/56147/hosting/docker-compose.yaml}} initializes couchdb with an initialization service. The \texttt{CKAN} project creates their own customer container that loads their database from a tabular file like CSV.\footnote{\url{https://github.com/ckan/ckan/blob/e1568fd/contrib/docker/docker-compose.yml}} The \texttt{graphql} engine project initializes their example project using a migration container.\footnote{\url{https://github.com/hasura/graphql-engine/blob/2325755/community/sample-apps/tic-tac-toe-react/docker-compose.yaml}}
    \item \textbf{Supporting Evidence.} These examples were found by counting the Docker Compose file that contained \textit{database service} and \textit{database init service} service types together as seen in \Cref{tab:service-types}.
\end{itemize}

\subsection{Database Administration Service with Database Service}
\label{pattern:data-admin}

\begin{figure}[htbp]
\centering
\includegraphics[width=0.5\linewidth]{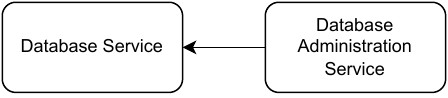}
  \caption{High-level overview of the Database Administration Service with Database Service pattern.}
\label{fig-pattern:data-admin}
\end{figure}

\begin{itemize}
    \item \textbf{Motivation.} Databases are often easier to manage with a graphical user interface (GUI).
    \item \textbf{Context/Applicability.} Developers wishing to access the database visually without installing any tools on their local machine and without using any external online web services.
    \item \textbf{Description.} A database administration service such as adminer or phpmyadmin are orchestrated into the same network as the database service. Users can then browse the database via the database administration service through a web browser. %
    \item \textbf{Advantages.} An orchestrated database can be managed with a graphical interface without any external concerns such as security.
    \item \textbf{Potential Issues.} The database administration service provides an interface into the database that can be exploited if the service is exposed.
    \item \textbf{Real-world Examples.} This pattern was discovered in 22 Docker Compose files.  The \texttt{libretime} project uses adminer to manage a postgres instance.\footnote{\url{https://github.com/LibreTime/libretime/blob/05342/docker-compose.yml}} The \texttt{personal management system} project uses adminer to manage a mariadb instance.\footnote{\url{https://github.com/Volmarg/personal-management-system/blob/8b377/docker-compose.yml}} The \texttt{unmark} project uses adminer to manage a mysql instance.\footnote{\url{https://github.com/cdevroe/unmark/blob/da271/docker-compose.yml}}
    \item \textbf{Supporting Evidence.} These examples were found by counting the service type combination of \textit{database service} and \textit{database administration service} present in Docker Compose files as seen in \Cref{tab:service-types}.
\end{itemize}

\subsection{Service Labels are used to Configure Reverse Proxy}
\label{pattern:service-labels-reverse}

\begin{figure}[htbp]
\centering
\includegraphics[width=0.9\linewidth]{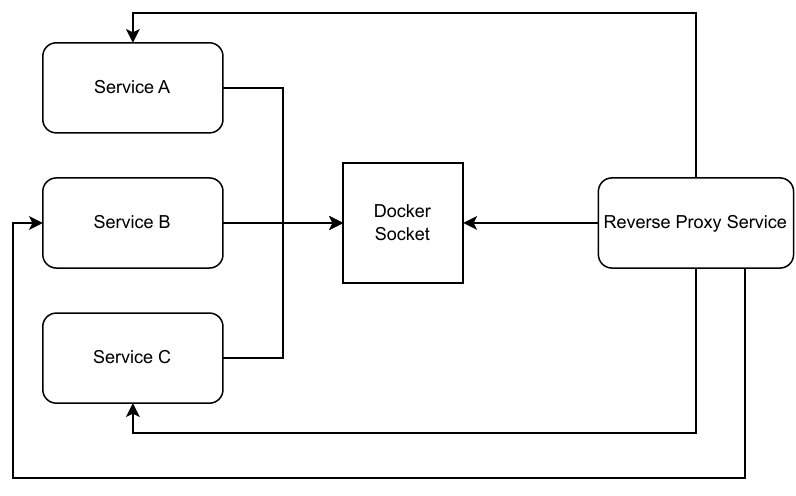}
  \caption{High-level overview of the Service Labels are used to Configure Reverse Proxy pattern.}
\label{fig-pattern:service-labels-reverse}
\end{figure}

\begin{itemize}
    \item \textbf{Motivation.} Reverse proxies such as nginx require a configuration file on how to route its proxy to different servers. Instead of using a configuration file, a reverse proxy can be configured with the Docker service labels. Docker service labels are labels that can be given to services in a Docker Compose file. The use of a reverse proxy as a pattern has been described by \citet{sousa2018overview,sousa2015patterns}.
    \item \textbf{Context/Applicability.} To simplify reverse proxy configuration, all reverse proxy configuration can be centralized to a Docker Compose file.
    \item \textbf{Description.} Services such as the \texttt{traefik} reverse proxy uses the Docker API (via the Docker Socket) to obtain the configuration of routing to services via their service labels. The \texttt{traefik} reverse proxy routes requests to the relevant containers. %
    \item \textbf{Advantages.} Instead of managing configuration files that need to be mounted to a reverse proxy service, the configuration can be centralized into a Docker Compose file.
    \item \textbf{Potential Issues.} It can be difficult to debug if the configuration for \texttt{traefik} does not work. There is an issue with Docker API security as the host may be able to be accessed via \texttt{traefik}'s access to the Docker API~\citep{traefik-docker-security}.
    \item \textbf{Real-world Examples.} We count 15 Docker Compose files using this pattern. The \texttt{isotope mail client project} uses labels to configure their \texttt{traefik} proxy.\footnote{\url{https://github.com/manusa/isotope-mail/blob/37722/deployment-examples/docker-compose.yml}} The \texttt{shaarli} bookmarking project uses labels to configure \texttt{traefik}.\footnote{\url{https://github.com/shaarli/Shaarli/blob/b7c50a5/docker-compose.yml}} The \texttt{zenbot} project has a Docker Compose file that uses \texttt{traefik} with services that have \texttt{traefik} service labels.\footnote{\url{https://github.com/DeviaVir/zenbot/blob/0cb3541/docker-compose-traefik.yml}}
    \item \textbf{Supporting Evidence.} These examples were found by counting the observations among services in \Cref{tab:holistic-service} under \qq{Service labels are configuration for reverse proxy}.
\end{itemize}

\subsection{Mail Service Testing}
\label{pattern:mail-testing}

\begin{figure}[htbp]
\centering
\includegraphics[width=0.5\linewidth]{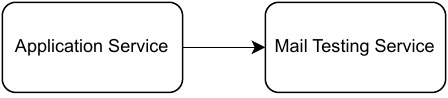}
  \caption{High-level overview of the Mail Service Testing pattern.}
\label{fig-pattern:mail-testing}
\end{figure}

\begin{itemize}
    \item \textbf{Motivation.} To test the email functionalities of an app, an SMTP service needs to be created. The creation of SMTP services can be complex hence the existence of SaaS providers like Mailtrap, SendGrid, Amazon SES, and Mailgun. 
    \item \textbf{Context/Applicability.} Developers wishing to keep development local may wish to orchestrate their own SMTP services.
    \item \textbf{Description.} When the mail service is orchestrated, the mail service may connect with the application service with environment variables explicitly. Otherwise the mail service is connected internally by being hard-coded into the application source code. %
    \item \textbf{Advantages.} There is no reliance on external services that require internet connections and mail can be retrieved locally as opposed to email providers that may block adhoc SMTP services that attempt to deliver to a provider's inbox.
    \item \textbf{Potential Issues.} Mail services are an additional service in the software system that have configuration and debugging costs. The alternative are SaaS providers which reduce these costs by providing SMTP credentials that are guaranteed to work with an Internet connection.
    \item \textbf{Real-world Examples.} This pattern was discovered in 62 Docker Compose files.  The \texttt{roadiz} project uses \texttt{mailhog} to test emails.\footnote{\url{https://github.com/roadiz/roadiz/blob/ac370b0/docker-compose.yml}} The \texttt{shopware} project also uses \texttt{mailhog} to test emails in development.\footnote{\url{https://github.com/shopware/platform/blob/d93ca0f/docker-compose.yaml}} The \texttt{umbraco CMS} project uses \texttt{smtp4dev} to test emails in development.\footnote{\url{https://github.com/umbraco/Umbraco-CMS/blob/a73c7bb/.devcontainer/docker-compose.yml}}
    \item \textbf{Supporting Evidence.} These examples were found by counting the service types in \Cref{tab:service-types} under \textit{mail service}.
\end{itemize}

\subsection{Application Service with Database}
\label{pattern:serv-database}

\begin{figure}[htbp]
\centering
\includegraphics[width=0.5\linewidth]{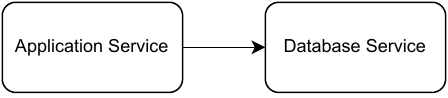}
  \caption{High-level overview of the Application Service with Database pattern.}
\label{fig-pattern:serv-database}
\end{figure}

\begin{itemize}
    \item \textbf{Motivation.} Most software systems contain some sort of backend database service. Therefore, it would be beneficial to know how these backend database services are orchestrated. 
    \item \textbf{Context/Applicability.} For any software system that does not store data in memory and stores data in a backend database service such as: MySQL, MariaDB, Postgres, or MongoDB. The backend service is not part of the main application container.
    \item \textbf{Description.} A main application uses the \qq{depends\_on} instruction to define a relationship between the main application and backend database service. This signifies that the main application will not start up until the backend database service has started. The backend database service mounts a volume for persistent storage, so that the data in the database is not destroyed when the \qq{docker compose down} command is run. The backend database service might also bind to a port on the host machine, the bound port can be used to directly access the database from the host machine. %
    \item \textbf{Advantages.} This pattern is useful when needing to orchestrate a backend service within your software system especially when testing and developing applications as containers are quite flexible. Spinning up and destroying databases simply requires the definition of a database service in a Docker Compose file.
    \item \textbf{Disadvantages.} If the backend database service is hosted on the same host machine, it can lead to catastrophic failures in a production environment when the host machine is down. 
    \item \textbf{Potential Issues.} Using \qq{depends\_on} does not guarantee that your application will connect to the backend database service on the first try. This is because although a container may have started, the service application might not be ready to accept applications. Therefore, \citet{docker-startup-order} recommends that a check is built into your application container to ensure the database service is ready to accept connections.
    \item \textbf{Real-world Examples.} We found 121 Docker Compose files using this pattern. For example, we can see this in the \texttt{accent} project\footnote{\url{https://github.com/mirego/accent/blob/22d5847/docker-compose.yml}}, \texttt{anchr} project\footnote{\url{https://github.com/muety/anchr/blob/320ecb5/docker-compose.yml}}, and \texttt{stringer} project\footnote{\url{https://github.com/swanson/stringer/blob/0b64dc0/docker-compose.yml}}.
    \item \textbf{Supporting Evidence.} This was found by counting the Docker Compose files that had the service types: \textit{frontend service} and \textit{database service} but not \textit{caching service} as seen in \Cref{tab:service-types}.
\end{itemize}

\subsection{Application Service with Database and Caching}
\label{pattern:serv-data-cache}

\begin{figure}[htbp]
\centering
\includegraphics[width=0.5\linewidth]{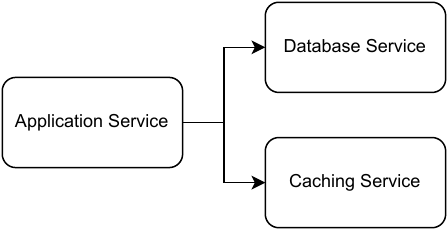}
  \caption{High-level overview of the Application Service with Database and Caching pattern.}
\label{fig-pattern:serv-data-cache}
\end{figure}

\begin{itemize}
    \item \textbf{Motivation.} To better support backend database services, often caching layers are used to improve performance when retrieving data from backend database services.
    \item \textbf{Context/Applicability.} This pattern is useful when looking to develop and debug software systems that have a caching mechanism. The most common caching service found is Redis.
    \item \textbf{Description.} The main application uses the \qq{depends\_on} instruction to define a relationship between the main application and backend database service. It uses another \qq{depends\_on} instruction to define a relationship between the caching service.
    The backend database service mounts a volume, so that the data in the database is not destroyed when the \qq{docker compose down} command is run. The caching service also mounts a volume too, to avoid data being ephemeral.
    The backend database service might also bind to a port on the host machine, the binded port can be used to directly access the database from the host machine. The caching service might also bind to a port on the host machine for direct access to the caching service. %
    \item \textbf{Advantages.} The pattern is useful for developing and testing software systems that wish to cache data. Since all the services are orchestrated locally, services can be easily accessed without any connectivity issues. 
    \item \textbf{Disadvantages.} Orchestrating all the services on the same host machine can lead to downtime as if any one service fails then the application may fail to run.
    \item \textbf{Potential Issues.} The backend database service and caching service containers may start up, but the services may not be ready to accept connections. As a result, \citet{docker-startup-order} recommends that the main application should implement a check to ensure that the database and caching services are online.
    \item \textbf{Real-world Examples.} This pattern was discovered in 42 Docker Compose files. For example, we can see this in the \texttt{miaou} project\footnote{\url{https://github.com/Canop/miaou/blob/7b251dc/docker-compose.yml}}, \texttt{wildduck} project\footnote{\url{https://github.com/nodemailer/wildduck/blob/ca3a365/docker-compose.yml}}, and \texttt{overleaf} project\footnote{\url{https://github.com/overleaf/overleaf/blob/fd36c4136/docker-compose.yml}}.
    \item \textbf{Supporting Evidence.} This was found by counting the Docker Compose files that had the service types: \textit{frontend service} and \textit{database service} and \textit{caching service} as seen in \Cref{tab:service-types}.
\end{itemize}

\subsection{HTTP Reverse Proxy Service}
\label{pattern:reverse-proxy}

\begin{figure}[htbp]
\centering
\includegraphics[width=0.6\linewidth]{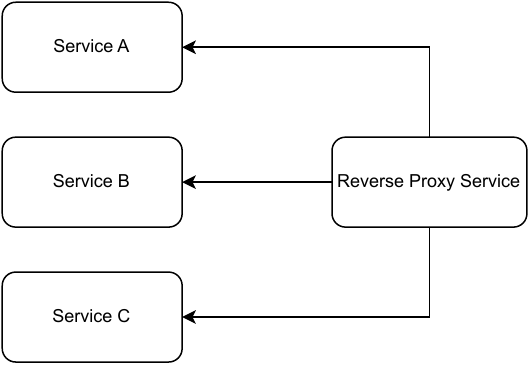}
  \caption{High-level overview of the HTTP Reverse Proxy Service pattern.}
\label{fig-pattern:reverse-proxy}
\end{figure}

\begin{itemize}
    \item \textbf{Motivation.} Services wishing to be exposed over a single port are combined using a reverse proxy service. \citet{sousa2018overview,sousa2015patterns,burns2016design} has described the use of a reverse proxy as a pattern. While \citet{fehling2014cloud} describes a similar \qq{Application Component Proxy} pattern.
    \item \textbf{Context/Applicability.} This pattern can be used when you wish to expose your service over default ports such as port 80 for HTTP and port 443 for HTTPS. Reverse proxy services such as Caddy and Nginx are often used as reverse proxies. 
    \item \textbf{Description.} This pattern works as follows:
    \begin{itemize}
        \item A reverse proxy service binds to ports such as 80 and 443. A volume is also mounted to the reverse proxy service to keep the configuration files persistent. The folder containing the log files can also be mounted to a volume.
        \item The reverse proxy service may use a \qq{depends\_on} instruction to define a relationship to the application(s) that it proxies. However, this is not necessary and can impede the startup time of accepting connections for the software system. 
    \end{itemize}
    \item \textbf{Advantages.} Managing the reverse proxy over a container reduces the need to install software on the host machine.
    \item \textbf{Disadvantages.} Since the reverse proxy is hosted within a container, there is additional overhead and complexity to running the reverse proxy software as the container needs to be managed. 
    \item \textbf{Potential Issues.} Using a reverse proxy over a container can be more difficult to debug when log files are not mounted to a volume.
    \item \textbf{Real-world Examples.} There are 92 Docker Compose files using this pattern. For example, we can see this in the \texttt{goodwork} project\footnote{\url{https://github.com/iluminar/goodwork/blob/f452b7d5/docker-compose.yml}}, \texttt{linkace} project\footnote{\url{https://github.com/Kovah/LinkAce/blob/6589c92/docker-compose.yml}}, and \texttt{PatrowlManager} project\footnote{\url{https://github.com/Patrowl/PatrowlManager/blob/7567c27/docker-compose.yml}}.
    \item \textbf{Supporting Evidence.} These examples were found by counting the service types in \Cref{tab:service-types} under \textit{reverse proxy service}.
\end{itemize}

\subsection{Duplicate Service Reuse}
\label{pattern:duplicate-service}

\begin{figure}[htbp]
\centering
\includegraphics[width=0.6\linewidth]{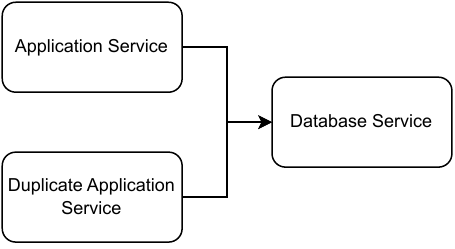}
  \caption{High-level example of the Duplicate Service Reuse pattern.}
\label{fig-pattern:duplicate-service}
\end{figure}

\begin{itemize}
    \item \textbf{Motivation.} There are two motivations for duplicate service reuse. First, applications sometimes bundle their functions into one image but would like their functions to be executed on separate instances. Second, application instances can be duplicated and isolated to test software systems that run multiple instances of the same kind of application in a \qq{dev} or \qq{test} mode.
    \item \textbf{Context/Applicability.} This pattern is useful when you wish to run multiple instances of your application to perform different functionalities of your application and to isolate these instances.
    \item \textbf{Description.} Services containing the same base image are defined in one of two ways. First, the service can be defined with multiple of the same image instances or be built from the same Dockerfile. Or the service can be duplicated using the \qq{YAML Anchor and Alias} pattern (\Cref{pattern:alias}) or \qq{Docker Compose Service Inheritance} pattern (\Cref{pattern:inheritance}). Next, the services are configured for their specific use cases, this might include changing the \qq{command} option or having different volumes and ports for the service. %
    \item \textbf{Advantages.} Multiple processes of the application can be easily managed. To update versions of containers only a single image needs to be pulled. There is reduced complexity in the amount of Docker images that need to be maintained.
    \item \textbf{Disadvantages.} Since the duplicated application can contain all functionalities of a software system, the codebase can be quite large making increasing the time to pull or build the Docker image of the duplicate application. Instead the application image could be separated into microservices.
    \item \textbf{Potential Issues.} If one part of the codebase is vulnerable in the application, the whole application could be compromised since they are all using the same image without separation of functionalities.
    \item \textbf{Real-world Examples.} We found 17 Docker Compose files using this pattern. For example, we can see this in the \texttt{mastodon} project\footnote{\url{https://github.com/tootsuite/mastodon/blob/8f7308da8/docker-compose.yml}}, \texttt{chaskiq} project\footnote{\url{https://github.com/chaskiq/chaskiq/blob/8fb33b/docker-compose.yml}}, and \texttt{Black Candy} project\footnote{\url{https://github.com/blackcandy-org/black_candy/blob/1d14fb9/docker-compose.yml}}.
    \item \textbf{Supporting Evidence.} These examples were found by counting the service types in \Cref{tab:service-types} under \qq{duplicate service reuse}.
\end{itemize}

\section{Discussion}
Our research is motivated by the popular usage of Docker Compose to deploy and manage software systems. We find that among configurations there are common kinds of service images used such as reverse proxies and databases. Furthermore, there is a common set of Docker Compose options used as well. This motivates us to investigate the patterns of configuration for the service images used to better understand their implications in self-hosted projects and wider applications.

\subsection{Identifying Common Patterns}
To identify common patterns we use an open coding process where we manually inspect each Docker Compose file and project to determine any interesting observations. As opposed to using strictly quantitative methods to discover patterns, we also use qualitative methods to perform an analysis. Using qualitative methods is beneficial as we can notice patterns that are not easily discoverable by statistics. For example, automatic Docker Compose file generation (\Cref{pattern:auto}) is a pattern that would not have been found from a quantitative analysis due to its low count. As well, the observation cannot be easily discovered by looking at Docker Compose files only because the project repository has to be inspected to understand how the files were created.

\subsection{Patterns in Self-hosted Projects}
Our patterns emerge from open coding software systems that are meant to be self-hosted on any infrastructure that one chooses. The patterns that we present in \Cref{sec:patterns} are representative of Docker Compose usage in software systems run by a diverse user base making it beneficial for developers wishing to use Docker Compose for their own projects. We find patterns that can help improve the developer experience such as using container management services described in \Cref{pattern:container-manage}. The patterns that we discover in self-hosted projects can be used as examples of how to orchestrate containers in future software systems.

\subsection{Patterns and Wider Applications}

Docker Compose users can benefit from these findings by better understanding patterns of what options to use when designing their software systems. %

This study helps to empirically validate what common patterns are already being used. Researchers of Docker can use the empirical observations of patterns of Docker Compose usage to motivate further studies on container orchestration in other development communities.  The empirical observations of this paper also suggest that there is merit to recovering architecture traces from Docker Compose files since containers that run services are often the building blocks of architecture (e.g.\ databases).

\subsection{Comparison with Other Infrastructure as Code Patterns}
\label{subsec:comparison}
We compare the patterns we discover with prior work discussed in \Cref{subsec:prior-work} and outline how our patterns are similar and different to demonstrate the practicality of our work. 

Prior research in \textit{Infrastructure as Code} (IaC), patterns of usage have been investigated by~\citep{rahman2019seven,ksontini2021refactorings,shamim2020xi,burns2016design, fehling2014cloud}. \citet{rahman2019seven} investigates security patterns in Puppet that deal with specific instructions that are used such as using hard-coded passwords. In comparison, our patterns focus on how to orchestrate services at a higher level. \citet{shamim2020xi} identify implementation practices that can better secure Kubernetes deployments such as implementing role-based access control authorization, continuously applying security patches, and implementing security policies for Kubernetes pods and networks. By contrast, our work focuses explicitly on how developers might deploy software services such as databases to solve typical software application deployments. 

\citet{burns2016design} identify patterns for container-based distributed systems. In our paper, we identify the \qq{HTTP Reverse Proxy Service} pattern (\Cref{pattern:reverse-proxy}) which resembles the ambassador pattern described by \citet{burns2016design} --- a proxy is used to route to different services from a single server. We also identify the \qq{Duplicate Service Reuse} (\Cref{pattern:duplicate-service}) that can be used for the multi-node application patterns described by \citet{burns2016design}. Therefore, certain patterns we use bear similarities to the patterns found in container-based distributed systems as described in \citep{burns2016design}.

\citet{ksontini2021refactorings} identify refactoring types for Dockerfiles and Docker Compose that suggest some best practices. Our patterns support many of their best practices identified from refactoring types. Our \qq{Docker Compose Override Use Case} pattern in \Cref{pattern:override} correspond with the refactoring types in \citep{ksontini2021refactorings}: \qq{Extract ports attribute into an override Docker-compose file}; \qq{Extract volume attribute into an override Docker-compose file}; \qq{Extract ENV attribute into an override Docker-compose file or use .env file}; \qq{Extract service into an override Docker-compose file}. While our \qq{Docker Compose Service Inheritance} pattern in \Cref{pattern:inheritance} corresponds with the \qq{Add Extends attribute to inherit configuration from an existing service thus avoiding duplication} refactoring type in \citep{ksontini2021refactorings}. 

While we do not identify patterns that explicitly set names for services, containers, and volumes, or adding tags to images like the refactoring types identified in \citep{ksontini2021refactorings}, we do identify the \qq{Container Management Services} pattern that can be used to maintain the latest images of services which negates the need for refactoring. We outline in our \qq{Application Service with Database} pattern in \Cref{pattern:serv-database} and \qq{Application Service with Database and Caching} in \Cref{pattern:serv-data-cache} how dependencies can be an issue in service orchestration which corresponds to the \qq{Order services based on their dependency order} refactoring type in \citep{ksontini2021refactorings}. Finally, \citet{ksontini2021refactorings} identify the \qq{Add ENV variables to store useful system-wide values} refactoring type, where we make suggestions in the \qq{Database Initialization Service with Database Service} pattern (\Cref{pattern:data-init}) and \qq{Mail Service Testing} pattern (\Cref{pattern:mail-testing}) to use environment variables to connect to these services. As a result, our patterns identified with our qualitative analysis align with many of the refactoring types identified by \citet{ksontini2021refactorings}.

\citet{sousa2018overview,sousa2015patterns} presents the idea of a reverse proxy pattern where services are routed to their appropriate network addresses. The most closely related pattern we find is the \qq{Service Labels are used to Configure Reverse Proxy} pattern described in \Cref{pattern:service-labels-reverse} where service routing can automatically be determined through service labels. Also related is using a reverse proxy in the \qq{HTTP Reverse Proxy Service} pattern (\Cref{pattern:reverse-proxy}).

\section{Threats to Validity}
In terms of construct validity, the projects that we analyze may not be representative of all real world usage of Docker Compose. As such, the empirical observations seen cannot be assumed to be applied to all projects using Docker Compose. Even so, the projects of the self-hosted community that we do analyze provides some nuance to usage that may not be obtained by analyzing large datasets with unknown contexts. Furthermore, we also assume that the git history of projects has not been rewritten. We also assume that the data is accurately retrieved and that the filtering process for identifying Docker Compose usage is accurate.

With regards to internal validity, we construct our patterns through an open coding process whereby patterns of usage are collected through an iterative process. Since we use an open coding process, some patterns may be overlooked due to the manual nature of the analysis. As well, there may be preconceived notions as to what patterns might be expected when performing the open coding process. To reduce the risk of preconceived notions and overlooking patterns, we review our data with an open mind in three dimensions: (1) at a service level in Docker Compose files, (2) at an orchestration level with how services in a Docker Compose configuration work together, and (3) at a repository level with how Docker Compose files operate within a project repository. 

Finally, in consideration of external validity, our dataset is not comprehensive and may not generalize to all projects. Since we only analyze publicly available projects, private projects such as internal company projects may use Docker Compose differently. Furthermore, full architectures may not be reflected since services can run on a single container (e.g., a database and application run in the same container). However, this is unlikely among most projects as it would defeat the purpose of using Docker Compose. We provide a replication package~\citep{replication-package} outlining our process and scripts for those wishing to analyze more projects.

\section{Conclusions and Future Work}

This paper qualitatively and quantitatively explored a set of successful self-hosted software-based service projects that employ Docker Compose to orchestrate and deploy various software services using Docker containers. Based on these analyses, typical patterns of service architecture, orchestration, and deployment were identified, counted, characterized, and named. Naming and characterizing the patterns allows others to discuss their architectures using the pattern names as shortcuts rapidly. These patterns also describe to developers patterns that have been successfully deployed. We demonstrate these projects are successful based on having more GitHub stars than prior studies like~\citep{ibrahim2021study}. The greater number of GitHub stars indicates a project's attention from a wider community and shows the value and significance of the project.

To encourage further patterns research and replicability, we contribute a replication package~\citep{replication-package} that includes an open-coded dataset of concepts in Docker Compose files, as well as scripts for analysis and extracting our selected Docker Compose projects.

We find many common patterns one would expect from self-hosted software services such as the use of an \qq{HTTP Reverse Proxy Service} (\Cref{pattern:reverse-proxy}), or \qq{Application Service with Database and Caching} (\Cref{pattern:serv-data-cache}). We found interesting patterns that have to do with deploying databases, especially initializing databases before the application can use them, as described by the \qq{Database Initialization Service with Database Service} pattern in \Cref{pattern:data-init}. The \qq{Service Labels are used to Configure Reverse Proxy} pattern described in \Cref{pattern:service-labels-reverse} is also interesting, as it shows how reverse proxies can be dynamically configured by Docker Compose leveraging service labels in the Docker Compose files used to deploy and orchestrate the app.

\section*{Data Availability}

The datasets generated and analyzed during this study are available as part of our replication package on \url{https://zenodo.org/records/10648448}.

\bibliographystyle{unsrtnat}
\bibliography{main}   %

\end{document}